\documentclass[%
preprint,
nofootinbib,
 amsmath,amssymb,
 aps,
prb,
]{revtex4-2}
\usepackage{xcolor,import}
\usepackage{graphicx}
\usepackage{dcolumn}
\usepackage{bm}

\begin{document}

\makeatletter
\let\@makefntextOrig\@makefntext
\def\@makefntext#1{\@makefntextOrig{\baselineskip=7.2pt #1}}
\makeatother


\title{Domain Wall Skyrmions in Chiral Magnets}

\author{Calum Ross}%
 \email{calum.ross@ucl.ac.uk}
\affiliation{Department of Mathematics, University College London, London WC1E 6BT, United Kingdom\\
and Department of Physics Research and Education Center for Natural Sciences, Keio
University, Hiyoshi 4-1-1, Yokohama, Kanagawa 223-8521, Japan
}%

\author{Muneto Nitta}
\email{nitta@phys-h.keio.ac.jp}
 \affiliation{Department of Physics $\&$ Research and Education Center for Natural Sciences,\\
 Keio University, Hiyoshi 4-1-1, Yokohama, Kanagawa 223-8521, Japan}


%

\date{\today}

\begin{abstract}

Domain wall skyrmions are skyrmions trapped inside a domain wall.
We investigate domain wall skyrmions in chiral magnets 
using a fully analytic approach.
Treating the Dzyaloshinskii-Moriya (DM) interaction perturbatively, 
 we construct the low-energy effective theory 
 of a magnetic domain wall 
in an $O(3)$ sigma model 
with the DM interaction  
and an easy-axis potential term, 
yielding  a sine-Gordon model. 
We then construct domain wall skyrmions as 
sine-Gordon solitons along the domain wall.
We also construct domain wall skyrmions 
on top of a pair of a domain wall and an anti-domain wall.
One of characteristic feature of domain wall skyrmions is 
that both skyrmions and anti-skyrmions are 
equally stable inside the domain wall, 
unlike the bulk in which only one of them is stable.

%
\end{abstract}

\maketitle


\newpage

\section{\label{sec:level1}Introduction}

Skyrmions are topological solitons proposed by Skyrme as 
a model of nuclei \cite{Skyrme:1962vh}. 
They have been shown to be the baryons of
large-$N_c$ quantum chromodynamics (QCD) \cite{Witten:1983tx}, 
and have been studied extensively \cite{MantonsBook,*multifaceted,
*Manton:2004tk}.
Recently, 
there has been great interest in magnetic skyrmions 
\cite{Bogdanov:1989,*Bogdanov:1995}, 
which are two-dimensional analogues of skyrmions in chiral magnets,
 from both fundamental and applied sciences, 
since they were 
realized in chiral magnets 
with the Dzyaloshinskii-Moriya (DM) interaction
\cite{Dzyaloshinskii,*Moriya:1960zz}
in 
 laboratory experiments  
\cite{doi:10.1126/science.1166767,
*doi:10.1038/nature09124,*doi:10.1038/nphys2045}
and have been proposed 
as 
 information carriers in ultradense memory and logic devices  
  with low energy consumption
\cite{doi:10.1038/nnano.2013.29}. 
In a certain parameter region of chiral magnets, 
a chiral soliton lattice is the ground state 
\cite{togawa2012chiral,*togawa2016symmetry,*KISHINE20151,
*PhysRevB.97.184303,*PhysRevB.65.064433,*Ross:2020orc}
where the energy of a single soliton is negative, and one dimensional modulated states have lower energy than skyrmions.
In another parameter region,
the ground state is a lattice of skyrmions 
in which the energy of a single skyrmion becomes negative 
\cite{Rossler:2006,
*Han:2010by,*Lin:2014ada,*Ross:2020hsw}. Finally, there are ferromagnetic regions of the phase diagram where skyrmions appear as positive energy solitons above the ferromagnetic ground state.
In addition, isolated skyrmions are also experimentally 
observable \cite{Romming:2013}.
Another recent development is observations of 
skyrmion tubes in 3D materials \cite{Wolf:2022}.

On the other hand, when there is an easy-axis potential term, 
magnetic domain walls have also been 
studied for a long time in particular for their application to magnetic memories 
\cite{doi:10.1126/science.1145799,KUMAR20221}.
It is thus natural to consider, 
by combining these two objects, 
one can expect 
that there should be 
potential applications  in constructing 
further useful nanodevices, such as the domain wall racetrack memory proposal \cite{doi:10.1126/science.1145799} which has been extended to a proposal using skyrmions \cite{tomasello2014strategy}.
As such a composite object, 
``domain wall skyrmions''
have been proposed 
in quantum field theory \cite{Nitta:2012xq,*Kobayashi:2013ju}\footnote{
Originally the term 
``domain wall skyrmions'' was first introduced in Ref.~\cite{Eto:2005cc} 
in which Yang-Mills instantons become 
3D skyrmions inside a domain wall.
} 
and have been recently 
 observed  experimentally in chiral magnets 
\cite{PhysRevB.102.094402,Nagase:2020imn,Yang:2021} (see also \cite{Kim:2017lsi}).
A first step at treating chiral magnetic domain walls theoretically is given in Refs.~\cite{PhysRevB.99.184412,KBRBSK}.

In quantum field theory,
domain wall skyrmions can be described by sine-Gordon kink configurations on top of a domain wall or anti-domain wall 
\cite{Nitta:2012xq,*Kobayashi:2013ju}:
when skyrmions are absorbed into a domain wall, 
they become sine-Gordon solitons 
(see Refs.~\cite{Jennings:2013aea,*Bychkov:2016cwc} 
for subsequent studies 
and Refs~\cite{
Sutcliffe:1992he,*Stratopoulos:1992hq,
*Kudryavtsev:1997nw,*Auzzi:2006ju} 
for earlier related works).
More precisely, a domain wall 
in an $O(3)$ sigma model with an easy-axis potential term 
$V=m^2(1-n_3^2)$
posseses
the moduli space (collective coordinates) 
 ${\mathbb R} \times S^1$ 
\cite{Abraham:1992vb,*Abraham:1992qv,*Arai:2002xa,*Arai:2003es}.  
Then, 
 the low-energy effective theory describing 
the low-energy dynamics of the domain wall can be  
constructed by the so-called moduli approximation 
\cite{Manton:1981mp,*Eto:2006uw}, 
yielding a nonlinear sigma model with the target space 
${\mathbb R} \times S^1$ in this case. 
When there is a second anisotropy term involving $n_1$, in addition to 
the aforementioned potential term $V$ 
in the $O(3)$ model, 
it induces a sine-Gordon potential 
on the $S^1$ part in the domain wall effective theory.
Then, the domain wall skyrmion can be expressed by a sine-Gordon soliton 
in the domain wall effective theory 
\cite{Nitta:2012xq}.
In the absence of the 
the second anisotropy $n_1$, 
the skyrmion is diluted along the domain wall 
and eventually disappears, 
once it is absorbed into the domain wall. 
In quantum field theory, domain wall skyrmions are generalized in various directions.\footnote{ 
Some generalizations in field theory are:
it was generalized to the ${\mathbb C}P^{N-1}$ model
 in Ref.~\cite{Fujimori:2016tmw}, 
where ${\mathbb C}P^{N-1} = SU(N)/[SU(N-1)\times U(1)]$ is 
a complex projective spece of complex dimension $N-1$.
The ${\mathbb C}P^{N-1}$ model also admits skyrmions 
due to $\pi_2[{\mathbb C}P^{N-1}] \simeq {\mathbb Z}$, 
and $N-1$ parallel domain walls 
in the presence of a suitable potential term as
a generalization of the easy-axis potential
\cite{Gauntlett:2000ib,*Tong:2002hi}.
Then, domain wall skyrmions in 
the ${\mathbb C}P^{N-1}$ model are 
$U(1)^{N-1}$ coupled sine-Gordon solitons inside 
${\mathbb C}P^{N-1}$ domain walls \cite{Fujimori:2016tmw}. 
It was also shown in Refs.~\cite{Nitta:2015mma,*Nitta:2015mxa} 
that skyrmions in the Grassmannian sigma model 
 become non-Abelian sine-Gordon solitons 
\cite{Nitta:2014rxa,*Eto:2015uqa}
inside a non-Abelian domain wall 
\cite{Eto:2005cc,Shifman:2003uh,*Eto:2008dm}. 
}
On the other hand,
originally skyrmions were proposed as a model of nuclei and
live in three spatial dimensions (3D)  \cite{Skyrme:1962vh}.
In this dimensionality,  
domain wall skyrmions are 
3D skyrmions absorbed into 
a domain wall, 
in which they become 2D skyrmions (lumps)  
 \cite{Nitta:2012wi,*Gudnason:2014nba,*Gudnason:2014hsa} 
 (see also Refs.~\cite{Kudryavtsev:1999zm,*Gudnason:2013qba,*Gudnason:2018oyx} for similar configurations).
 Higher-dimensional domain wall skyrmions were further proposed 
 \cite{Nitta:2012rq}.
  Such domain wall skyrmions play a key role 
  to understand relations between topological solitons in quantum field theory 
  as proposed in Refs.~\cite{Nitta:2013cn,*Nitta:2013vaa,*Nitta:2022ahj}.

In this paper, we study, in a fully analytic way,  domain wall skyrmions 
in chiral magnets 
described by an $O(3)$ sigma model 
with a DM interaction  
and an easy-axis potential term.
Working in a small coupling regime for the DM term, 
we construct a domain wall effective theory 
and find that it is a sine-Gordon model 
with the DM term supplying the potential, 
even in the absence of the second anisotropy term for $n_1$.
We then construct domain wall skyrmions as 
sine-Gordon solitons  
in terms of the domain wall effective theory.
We also construct domain wall skyrmions 
on top of a pair of a domain wall and an anti-domain wall, 
and a domain wall skyrmion lattice.
We find that
that both skyrmions and anti-skyrmions are 
equally stabe inside the domain wall, 
in contruct to the bulk where only either of them is stable.
The analytic method proposed in this paper 
should be useful for further studies of domain wall skyrmions 
and proposing possible applications to nanotechnology.

This paper is organized as follows.
In Sec.~\ref{sec:model}, we give our model, and discuss 
the Derrick's scaling argument \cite{Derrick:1964} for stability of topological solitons in chiral magnets.
In Sec.~\ref{sec:effective energy}, we construct the low-energy effective action of a single domain wall and present the domain wall skyrmion solutions as well as a configuration where a lattice of sine-Gordon kinks is placed on top of a domain wall. 
Section \ref{sec:conclusions} is devoted to a summary and discussion.


\section{\label{sec:model}The Model}
In this section, we describe the model that we consider in this paper and 
apply the Derrick's scaling argument to chiral magnets.

\subsection{\label{sec:background}Energy functionals}
The static energy of a chiral magnet is described by an energy functional in terms of the magnetisation vector field 
$\vec{n} = (n_1,n_2,n_3):\mathbb{R}^{2}\to S^{2}$, a normalised three vector. Considering a specific model where there is both a DM term and 
an easy-axis anisotropy term in the $z$ direction leads to 
an $O(3)$ model:
\begin{equation}
\mathcal{E}=\frac{1}{2}\nabla\vec{n}\cdot\nabla\vec{n}+\kappa\,\vec{n}\cdot\left(\nabla_{-\alpha}\times\vec{n}\right)+m^{2}\left(1-n_{3}^{2}\right).
\end{equation}
The rotated gradient $\nabla_{-\alpha}$ with $\alpha\in(0,2\pi]$ shows that we have a one parameter family of models as discussed in Ref.~\cite{BRS}. Changing $\alpha$ changes the type of DM term with $\alpha=0$ corresponding to a Bloch DM term and $\alpha=\frac{\pi}{2}$ a N\'{e}el DM term. 

A convenient representation of the energy is using a complex stereographic coordinate $u \in {\mathbb C}$ related to the magnetisation vector field through $u=\frac{n_{1}+in_{2}}{1+n_{3}}$. In this coordinate, 
the static energy can be rewritten in the form of the ${\mathbb C}P^1$ model as
\begin{eqnarray}
&& E[u] = 2\int d^2x \left[\frac{
   \left\vert\nabla u\right\vert^{2}
+ 2\kappa\text{Im}\left(e^{i\alpha}\left[\partial_{z}u+u^{2}\partial_{z}\bar{u}\right]\right)
+ 2m^{2}\left\vert u\right\vert^{2}
}{\left(1+\vert u\vert^{2}\right)^{2}}\right]=E_{H}+E_{DM}+E_{0},\label{eq:chiral magnetic energy}\\
&&\quad E_{H} =  2\int d^2x \frac{\left\vert\nabla u\right\vert^{2}}{\left(1+\vert u\vert^{2}\right)^{2}} ,\\ 
&&\quad E_{DM} =  2\int d^2x \frac{  2\kappa\text{Im}\left(e^{i\alpha}\left[\partial_{z}u+u^{2}\partial_{z}\bar{u}\right]\right)   }{\left(1+\vert u\vert^{2}\right)^{2}}, \label{eq:DM}\\
&& \quad E_{0} = 2\int d^2x \frac{2m^{2}\left\vert u\right\vert^{2}}{\left(1+\vert u\vert^{2}\right)^{2}},
\end{eqnarray}
where $E_{H}$ is the Heisenberg, or Dirichlet, term in the energy involving two derivatives, $E_{DM}$ is the DM interaction term, and $E_{0}$ is the anisotropy term.

In terms of minimising the energy, it is well known that the three terms in Eq.~\eqref{eq:chiral magnetic energy} all want different things: The Heisenberg term wants the magnetisation vector field to align at nearby points, the potential term is easy axis anisotropy so wants the magnetisation to point in the third direction $\vec{n}=\pm \hat{e}_{3}$, finally the DM term wants the magnetisation vector to twist with the nature of the twist determined by the material parameter $\alpha$. In particular as pointed out in Ref.~\cite{BRS} when $\alpha=0$ the DM is of Bloch type and describes materials supporting Bloch type skyrmions, while $\alpha=\frac{\pi}{2}$ gives a N\'{e}el type DM term and describes materials supporting N\'{e}el type skyrmions.

\subsection{\label{sec:stability} Stability/ Derrick Scaling} Given a static energy functional such as Eq.~\eqref{eq:chiral magnetic energy}, it is natural to ask if energy minimising solutions exist. The standard approach is to use a Derrick scaling argument \cite{Derrick:1964,Manton:2004tk} to show that static solutions exist. This scaling argument is naturally dimension dependent and here we want to keep track of both the two dimensional and one dimensional case, 
relevant for skyrmions and domain walls, respectively.

Let $u_{\lambda}=u(\lambda x)$ be a scaled field and then treat the $1$D and $2$D cases separately.
\begin{itemize}
\item[1D:] In one dimension, the DM term is scale invariant since it involves one integral and one derivative, while the potential and Heisenberg terms scale oppositely to each other,
\begin{equation}
E[u_{\lambda}]=\lambda E_{H}[u]+E_{DM}[u]+\lambda^{-1}E_{0}[u].
\end{equation}
There are stable static solutions if this has a stationary point as a function of $\lambda$, in other words when $\left.\frac{d E}{d\lambda}\right\vert_{\lambda=1}=0$. This happens when
\begin{equation}
E_{H}[u]=E_{0}[u].
\end{equation}
Thus, the DM term is not needed for stable domain wall solutions to exist.

\item[2D:] In this case the Heisenberg term is scale invariant and the energy scales as
\begin{equation}
E[u_{\lambda}]=E_{H}[u]+\lambda^{-1}E_{DM}[u]+\lambda^{-2}E_{0}[u].
\end{equation}
Here $\left.\frac{d E}{d\lambda}\right\vert_{\lambda=1}=0$ implies that
\begin{equation}
E_{DM}=-2E_{0},
\end{equation}
which is possible since the DM term can give a negative contribution to the energy. Due to the chirality of the DM term, for a given material either skyrmions or anti-skyrmions will have negative DM energy. The other will have positive DM energy and is thus unstable. So either skyrmions or anti-skyrmions are stable in a given material but the other are unstable.\footnote{
An exception to this is the solvable model of Ref.~\cite{BRS} where both skyrmion, $Q=-1$, and anti-skyrmion $Q=1$ configurations are explicitly constructed. The resolution is that in the solvable model the $Q=1$ configurations are not true anti-skyrmions, but rather a superposition of a skyrmion and two anti-skyrmions which is stable.}
\end{itemize}

\section{\label{sec:effective energy}Domain wall skyrmions from low-energy effective theory
}

\subsection{\label{sec:Dw sols}Domain wall solutions in the absence of the DM tern}
From Refs.~\cite{Abraham:1992vb,*Abraham:1992qv,*Arai:2002xa,*Arai:2003es,
Nitta:2012xq,*Kobayashi:2013ju} 
it is known that in the absence of the DM term there are static domain wall solutions.  Using the stereographic coordinate $u$, 
the domain wall and anti-domain wall configurations are
\begin{equation}
u_{dw}=e^{ m\left(x_{1}-X\right)+i\varphi}, \qquad u_{adw}=e^{ -m\left(x_{1}-X\right)+i\varphi}, \label{eq:domain wall - anti-domain wall profile}
\end{equation}
with $X,\varphi$ moduli parameters corresponding to translation and phase of the wall, respectively. 
Both of these solutions have the same energy when $\kappa=0$, the energy is the domain wall tension
\begin{equation}
E[u_{dw}]=\vert T\vert=\left\vert\frac{m}{2}\left[\frac{1-\vert u\vert^{2}}{1+\vert u\vert^{2}}\right]^{x=+\infty}_{x=-\infty}\right\vert = m.
\end{equation}

By using the moduli approximation 
\cite{Manton:1981mp,*Eto:2006uw}, 
we promote the moduli $X,\varphi$ fields on the wall to be functions of the coordinates orthogonal to the wall, for our two dimensional model this is the $x_{2}$ direction.
From Refs.~\cite{Nitta:2012xq,Kobayashi:2013ju}, the effective energy along the domain wall for the Heisenberg and anisotropy terms is
\begin{equation}
E_{\text{eff}}=\frac{1}{2m}\left(m^{2}\partial_{2}X\partial^{2}X +\partial_{2}\varphi\partial^{2}\varphi\right),
\end{equation}
where a constant term $-m$ has been subtracted from the effective energy. 

\subsection{\label{sec:effective energy small kappa}Effective energy for small $\kappa$ and kinks solutions}
Turning on a small DM term perturbatively, $\kappa<1$, we can still take the domain wall and anti-domain wall solutions as valid configurations and we can consider fluctuations in $X$ and $\varphi$ around the domain wall and anti-domain wall solutions. 
Substituting in the (anti-) domain wall profile from Eq.~\eqref{eq:domain wall - anti-domain wall profile} 
to the DM term in Eq.~(\ref{eq:DM}), it becomes
\begin{equation}
\begin{split}
E_{DM}[u]=&\int_{-\infty}^{\infty}dx^{1}\left[\frac{2\kappa}{\left(1+\vert u\vert^{2}\right)^{2}}\vert u\vert \left(\pm m\sin\left(\varphi+\alpha\right)\mp m\partial_{2}X\cos\left(\varphi+\alpha\right)-\frac{d}{dx_{2}}\left(\cos\left(\varphi+\alpha\right)\right)\right)\right.\\
			&\left.+\frac{2\kappa}{\left(1+\vert u\vert^{2}\right)^{2}}\vert u\vert^{3} \left(\pm m\sin\left(\varphi+\alpha\right)\mp m\partial_{2}X\cos\left(\varphi+\alpha\right)+\frac{d}{dx_{2}}\left(\cos\left(\varphi+\alpha\right)\right)\right) \right].
\end{split}
\end{equation}
The two integrals are the same
\begin{equation}
\int_{-\infty}^{\infty}\frac{\vert u\vert^{3}}{\left(1+\vert u\vert^{2}\right)^{2}}dx^{1}=\frac{\pi}{4m}=\int_{-\infty}^{\infty}\frac{\vert u\vert}{\left(1+\vert u\vert^{2}\right)^{2}}dx^{1},
\end{equation}
which means that the boundary terms in the effective energy cancel and the contribution due to the DM term is
\begin{equation}
E_{DM}=\pm\pi\kappa\left[\sin\left(\varphi+\alpha\right)-\left(\partial_{2}X\right)\cos\left(\varphi+\alpha\right)\right].
\end{equation}
Adding this to the known effective energy from \cite{Nitta:2012xq,*Kobayashi:2013ju} gives
\begin{equation}
E_{\text{eff}}=\frac{1}{2m}\left(m^{2}\partial_{2}X\partial^{2}X+\partial_{2}\varphi\partial^{2}\varphi \pm\tilde{\kappa}\left[\sin\left(\varphi+\alpha\right)-\partial_{2}X\cos\left(\varphi+\alpha\right)\right]\right)
\end{equation}
with $\tilde{\kappa}=2m\pi\kappa$. 

This derivation is for a domain wall perpendicular to the $x^{1}$ direction. However,  if the domain wall is instead perpendicular to $\tilde{x}=x^{1}\cos\theta+x^{2}\cos\theta$ then the (anti-) domain wall solution becomes
\begin{equation}
u=e^{\pm m\left(\tilde{x}-X\right)+i\varphi},
\end{equation}
and the only term which changes is the DM term where $\varphi\to \varphi+\theta$.
Thus the effective energy density along the (anti-)domain wall is
\begin{equation}
\mathcal{E}[u_{dw}]=\frac{1}{2m}\left[\left(\partial_{2}X\right)^{2}+\left(\partial_{2}\varphi\right)^{2}\pm\tilde{\kappa}\sin\left(\alpha+\theta+\varphi\right)\pm2\frac{\tilde{\kappa}}{m}\partial_{2}X\cos\left(\alpha+\theta+\varphi\right)\right]. \label{eq:dw effective energy}
\end{equation}

When the translation modulus is constant, $\partial_{2}X=0$, the domain wall is straight, and this reduces to the sine-Gordon model, 
\begin{equation}
\mathcal{E}[u_{dw}]=\frac{1}{2m}\left[\left(\partial_{2}\varphi\right)^{2}\pm\tilde{\kappa}\sin\left(\alpha+\theta+\varphi\right)\right]
\end{equation}
with potential term $U(\varphi)=\pm\tilde{\kappa}\sin\left(\alpha+\theta+\varphi\right)$. 

We said above that the value of $\alpha$ dictates the type of DM term. It also differentiates between Bloch and N\'{e}el type domain walls, this is because the value of $\alpha$ determines the ground state of the effective energy in Eq.~\eqref{eq:dw effective energy}.

\begin{figure}[!htbp]
 \centering
\includegraphics[width=0.6\textwidth]{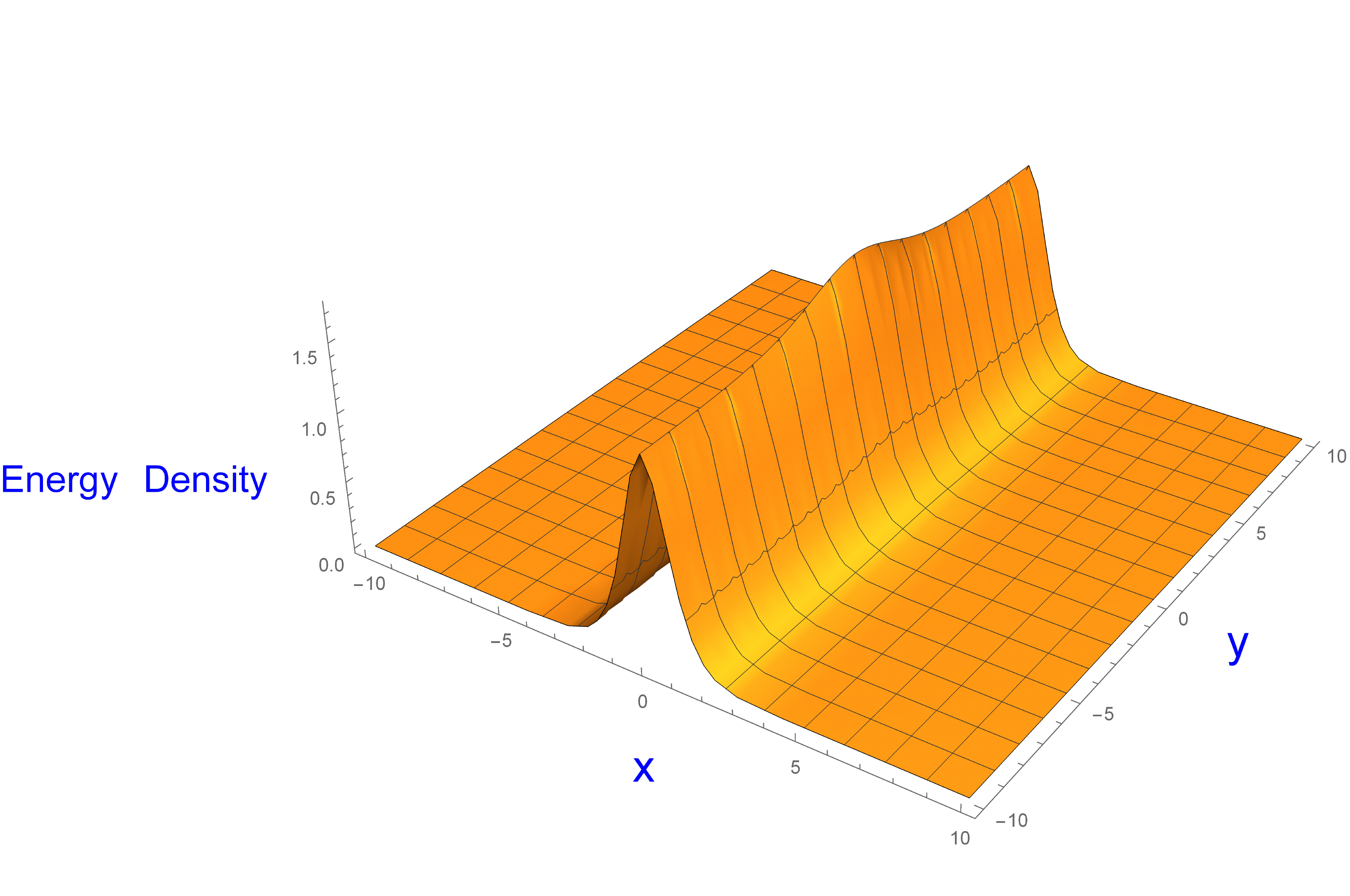}
\includegraphics[width=0.29\textwidth]{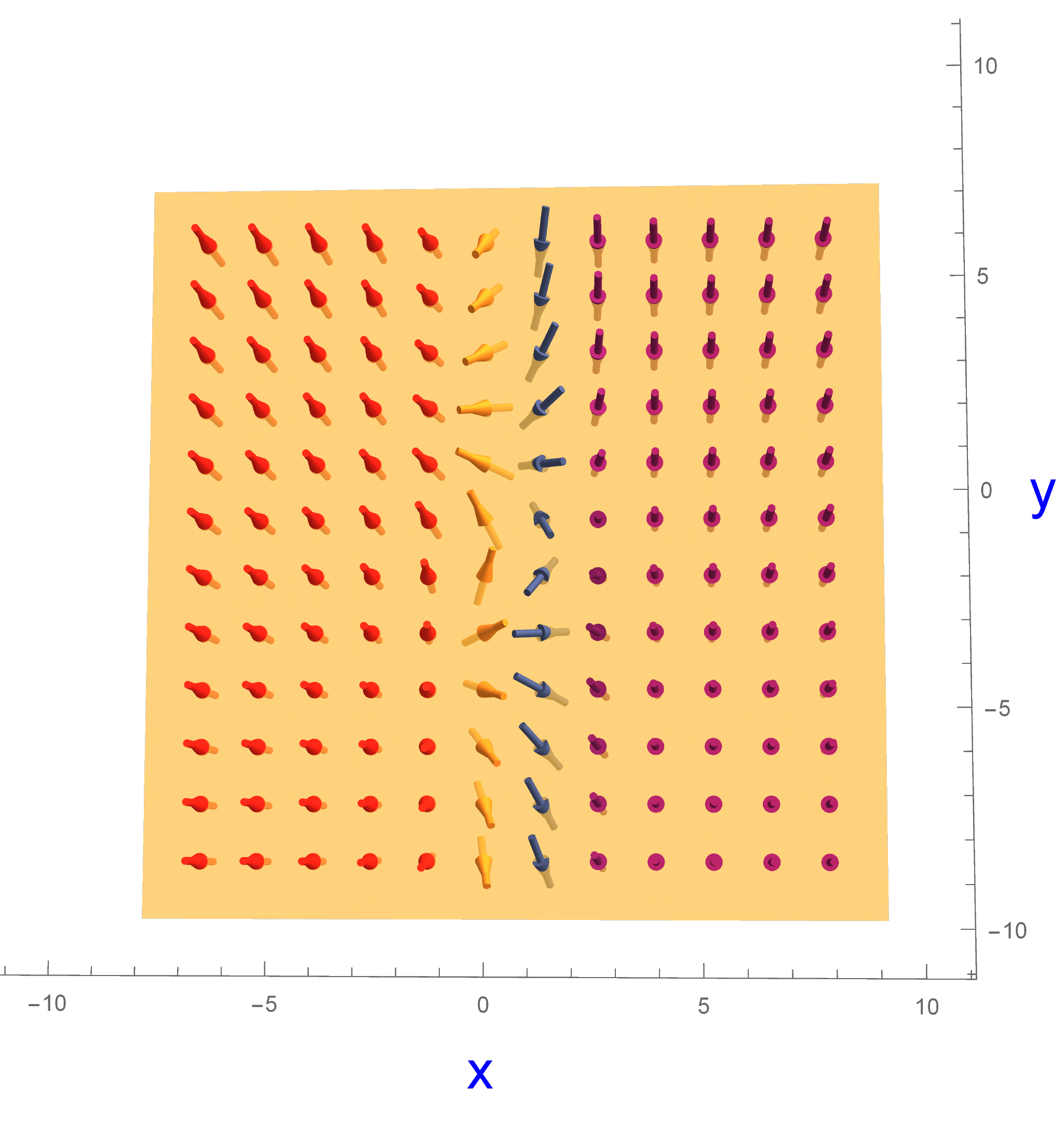}
 \caption{The energy density and magnetisation plots for a domain wall with a kink on it. For simplicity we have chosen $\tilde{\kappa}=\frac{1}{5},\alpha=\theta=X=c=0$ to give a kink centred at zero on the domain wall.}
 \label{fig:dw kink plots}
\end{figure}

The kink solutions are found by solving the first order equation 
\begin{equation}
\frac{d\varphi}{dx^{2}}=\pm\sqrt{U(\varphi)+C_{0}} \label{eq:Bogomol'nyi equations}
\end{equation}
when $C_{0}=\tilde{\kappa}$ is the first integral of the system, if we think of $x_{2}$ as the analogue of the time coordinate. A derivation of this equation and its solutions is given in Appendix.~\ref{appendix:kink solutions}. The kink and anti-kink solutions are phase shifted versions of the familiar sine-Gordon kinks:
\begin{equation}
\varphi=4\arctan\left(e^{\pm\sqrt{\frac{\tilde{\kappa}}{2}}x^{2}}\right)-\alpha-\theta-\frac{\pi}{2} 
\label{eq:sine-gordon kink and anti kink on domain wall}
\end{equation}
on $u_{dw}$, and 
\begin{equation}
\varphi=4\arctan\left(e^{\pm\sqrt{\frac{\tilde{\kappa}}{2}}x^{2}}\right)-\alpha-\theta+\frac{\pi}{2} 
\label{eq:sine-gordon kink and anti kink on anti-domain wall}
\end{equation}
on $u_{adw}$. The energy density and magnetisation for a kink on a domain wall are shown in Fig.~\ref{fig:dw kink plots}.

The topological charges (skyrmion numbers) are given in Table.~\ref{table:dwskyrme charges}. The topological charges are computed from 
\begin{equation}
Q[u]=\frac{i}{2\pi}\int \frac{\partial_{1}u\partial_{2}\bar{u}-\partial_{2}u\partial_{1}\bar{u}}{\left(1+\vert u\vert^{2}\right)^{2}}d^{2}x,
\end{equation}
for a domain wall skyrmion this becomes
\begin{equation}
Q[u]=\begin{cases}
\phantom{-}k\qquad \text{for a domain wall,}\\
-k\qquad \text{for an anti-domain wall,}
\end{cases}
\end{equation}
with $k$ the winding number of the sine-Gordon kink.
\begin{table}[!htbp]
\begin{tabular}{|l|l|l|}
\hline
                 & \textbf{Kink} & \textbf{Anti-Kink} \\ \hline
\textbf{Domain Wall}      & $Q=1$           & $Q=-1$               \\ \hline
\textbf{Anti-Domain Wall} & $Q=-1$          & $Q=1$                \\ \hline
\end{tabular}
\caption{The topological charges for the four basic combinations of domain wall skyrmions.}
\label{table:dwskyrme charges}
\end{table}
Since in the chiral magnets literature, the $Q=-1$ configuration is known as a skyrmion we should call an anti-kink on a domain wall a domain wall skyrmion.

One of the remarkable features of the domain wall skyrmions is that 
both skyrmion and anti-skyrmion are stable on the domain wall, 
unlike in the bulk where only either skyrmions or anti-skyrmions 
are stable 
and the others are unstable. 
This fact can be understood from the Derrick's scaling argument since both the domain wall and kink are separately stable configurations for their respective energies.

\begin{figure}[!htbp]
 \centering
\includegraphics[width=0.6\textwidth]{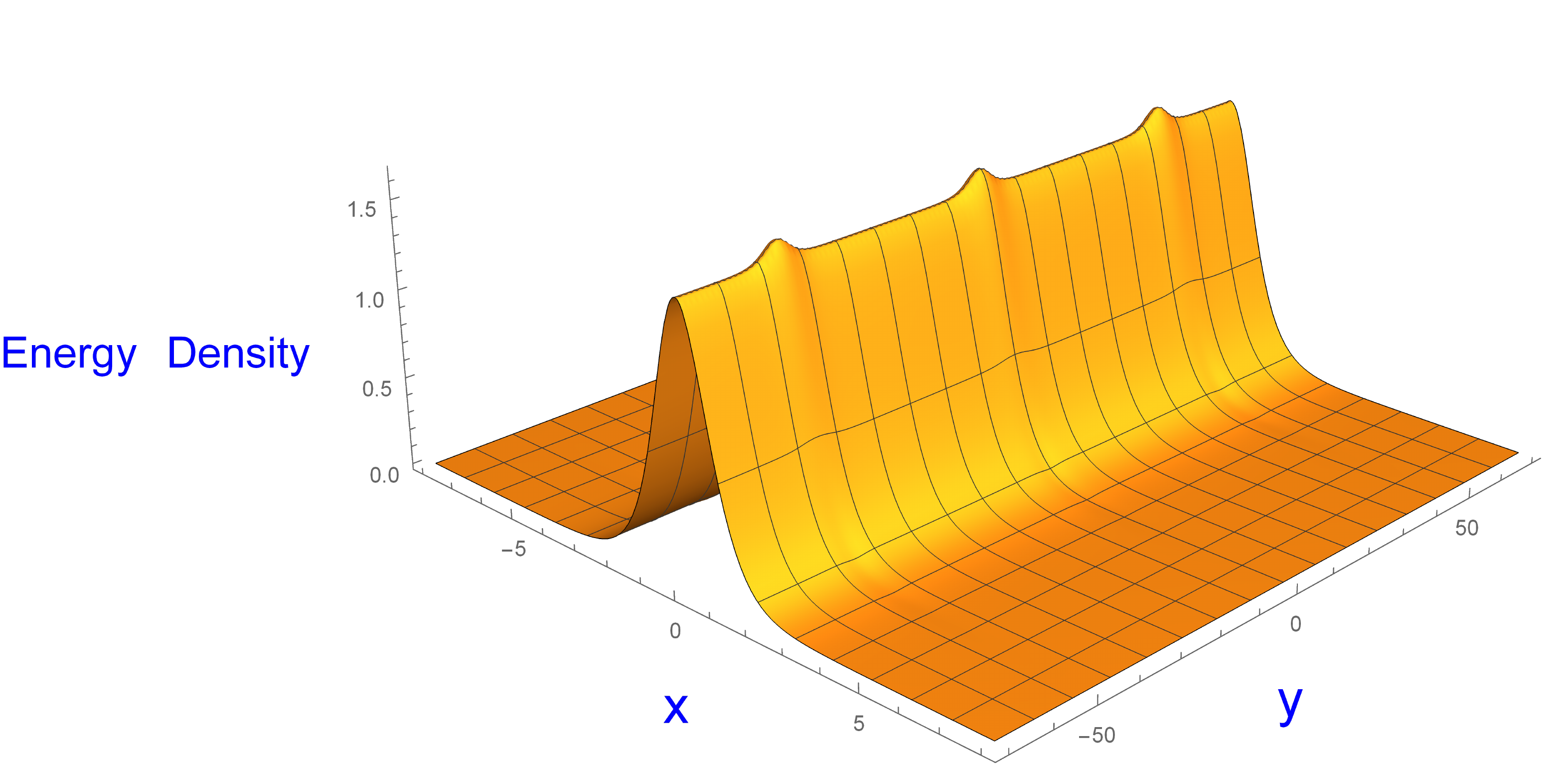}
\includegraphics[width=0.29\textwidth]{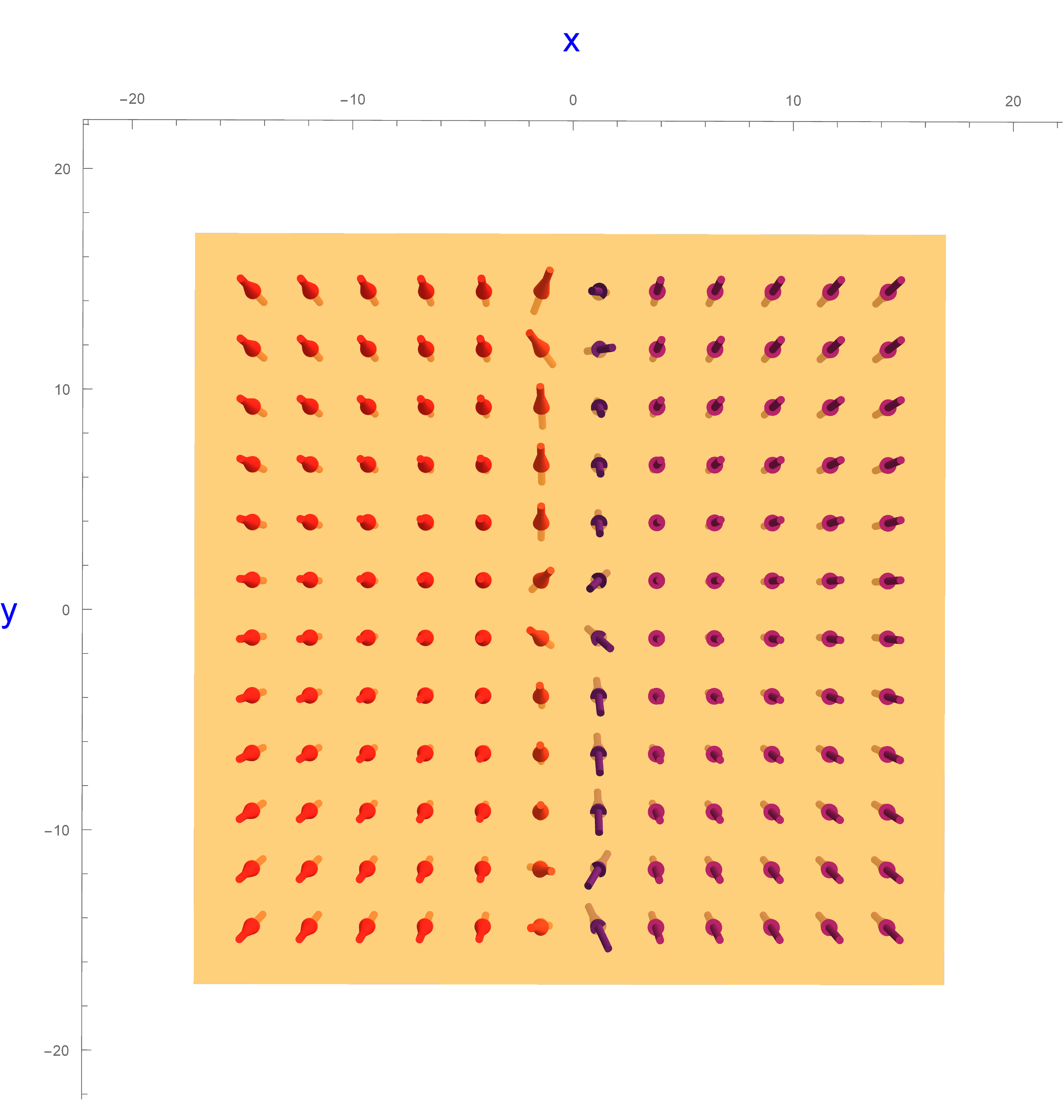}
 \caption{On the left, the energy density of a spiral configuration with $\tilde{\kappa}=\frac{1}{5}$. On the right is a magnetisation plot of the $k=\frac{99998}{100000}$ spiral on a domain wall. Comparing the the magnetisation plot in Fig.~\ref{fig:dw kink plots} we see that the spiral involves multiple kinks along the domain wall. }
 \label{fig:dw multikink plots}
\end{figure}

More generally we can consider higher energy solutions with a spiral on a domain wall or anti-domain wall. The derivation of these spiral solutions is given in Appendix.~\ref{appendix:multikink solutions}, the energy density for a spiral on a domain wall and a plot of the corresponding magnetisation is given in Fig.~\ref{fig:dw multikink plots}. The spiral configurations are described in terms of the elliptic modulus $k^{2}=\frac{2\tilde{\kappa}}{C_{0}+\tilde{\kappa}}$, where $C_{0}$ is the conserved first integral of Eq.~\eqref{eq:first integral}. These spirals on domain wall configurations have topological charge $Q=\pm k$ where $k$ is the number of domain walls in the spiral, and have a higher energy than the single domain wall skyrmion. These spiral configurations are sometimes known as the chiral soliton lattice \cite{KISHINE20151}, since they can be viewed as a lattice configuration of sine-Gordon kinks.

\subsection{\label{sec:interactions}Superposition configurations}
For ``well-separated'' domain wall anti-domain wall pairs, 
we can consider the configuration 
\begin{equation}
u_{W-aW}=e^{-m\left(x-X^{1}\right)+i\varphi_{1}}+e^{m\left(x-X^{2}\right)+i\varphi_{2}}. \label{eq:wall-anti-wall superposition}
\end{equation}
The case of constant phases $\varphi_{1}=\varphi_{2}+\pi$ was considered in the absence of a DM term in 
Ref.~\cite{Nitta:2012kj,*Nitta:2012kk,*Nitta:2012mg}. 
For well separated domain walls, $\vert X^{2}-X^{1}\vert \gg \frac{1}{m}$, this superposition is a valid configuration, assuming also that $X^{2}>X^{1}$.

\begin{figure}[!htbp]
 \centering
\includegraphics[width=0.6\textwidth]{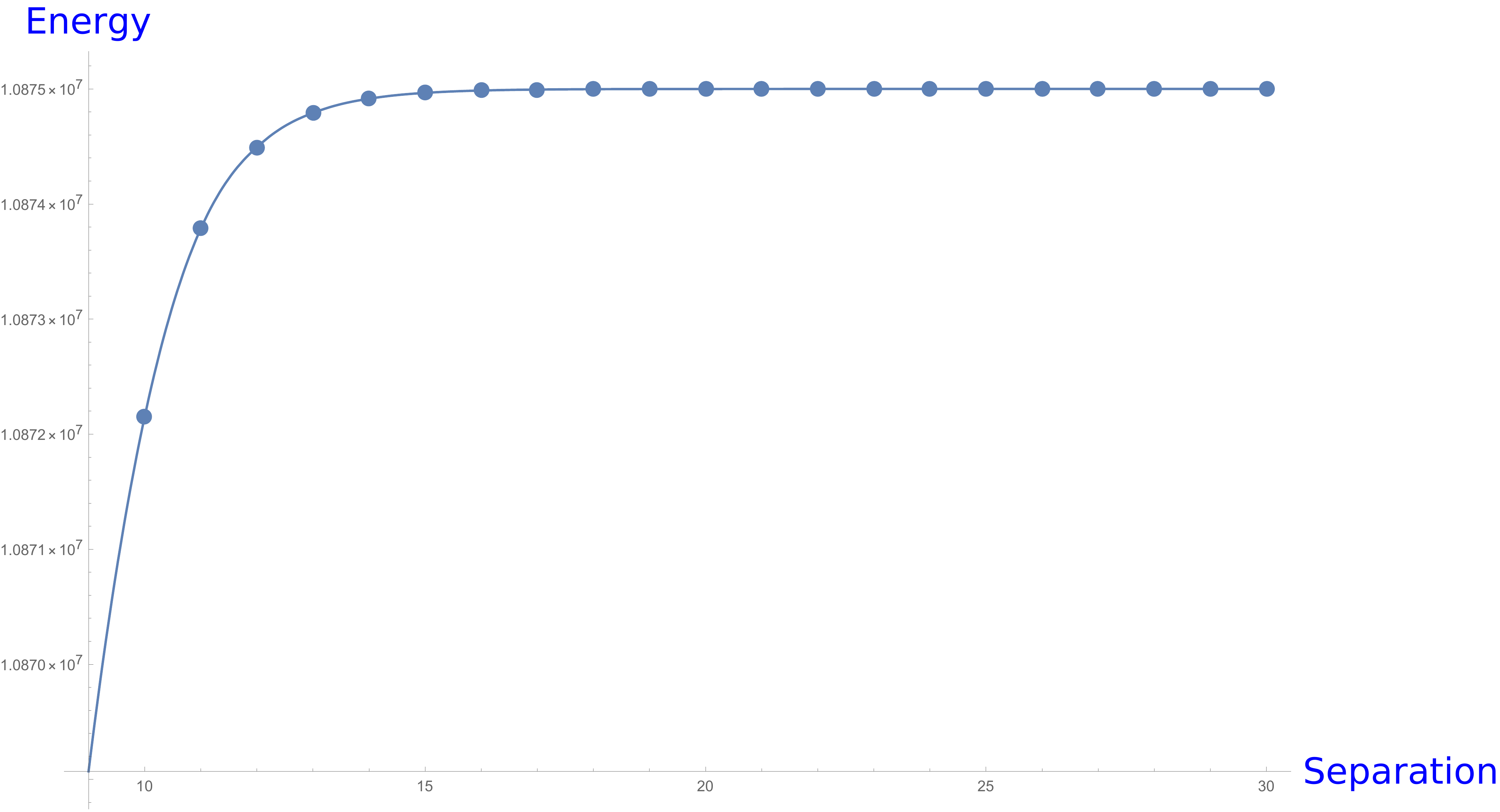}
 \caption{The energy for the superposition against the domain wall separation for the case of $m=1$.}
 \label{fig:dw_interaction energy}
\end{figure}

There are two separations that we can vary; one is $R=X^{2}-X^{1}$ the separation between the domain wall and the anti-domain wall, the other is when $\varphi_{i}$ is a kink or anti-kink we can vary the $x^{2}$ separation of the kink positions. 
Computing how the energy of this superposition varies with the separation of the domain wall and the anti-domain wall we see that this replicates the expected result that the domain wall and anti-domain wall have an attractive interaction. See Fig.~\ref{fig:dw_interaction energy} for a plot of the energy of the superposition against the domain wall separation $R$.

\begin{figure}[!htbp]
 \centering
\includegraphics[width=0.7\textwidth]{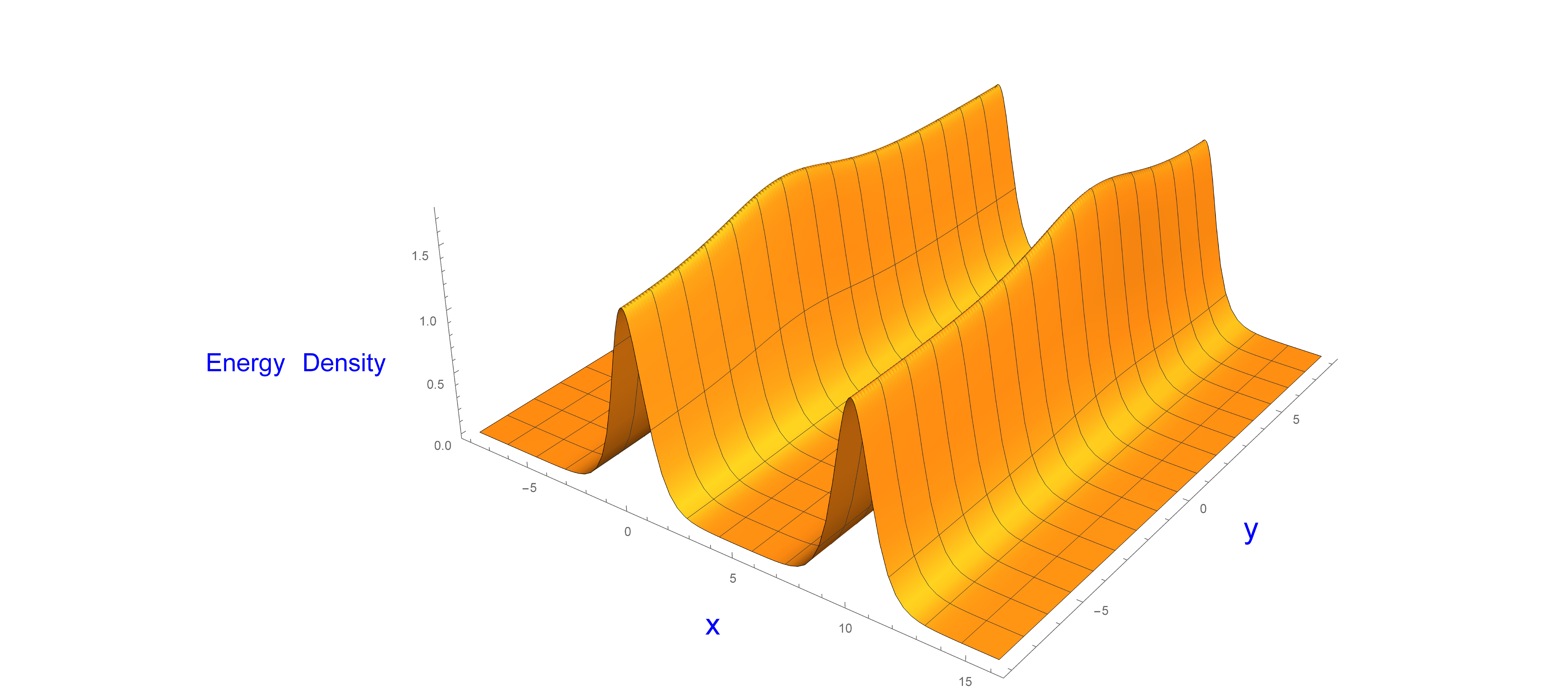}
\includegraphics[width=0.29\textwidth]{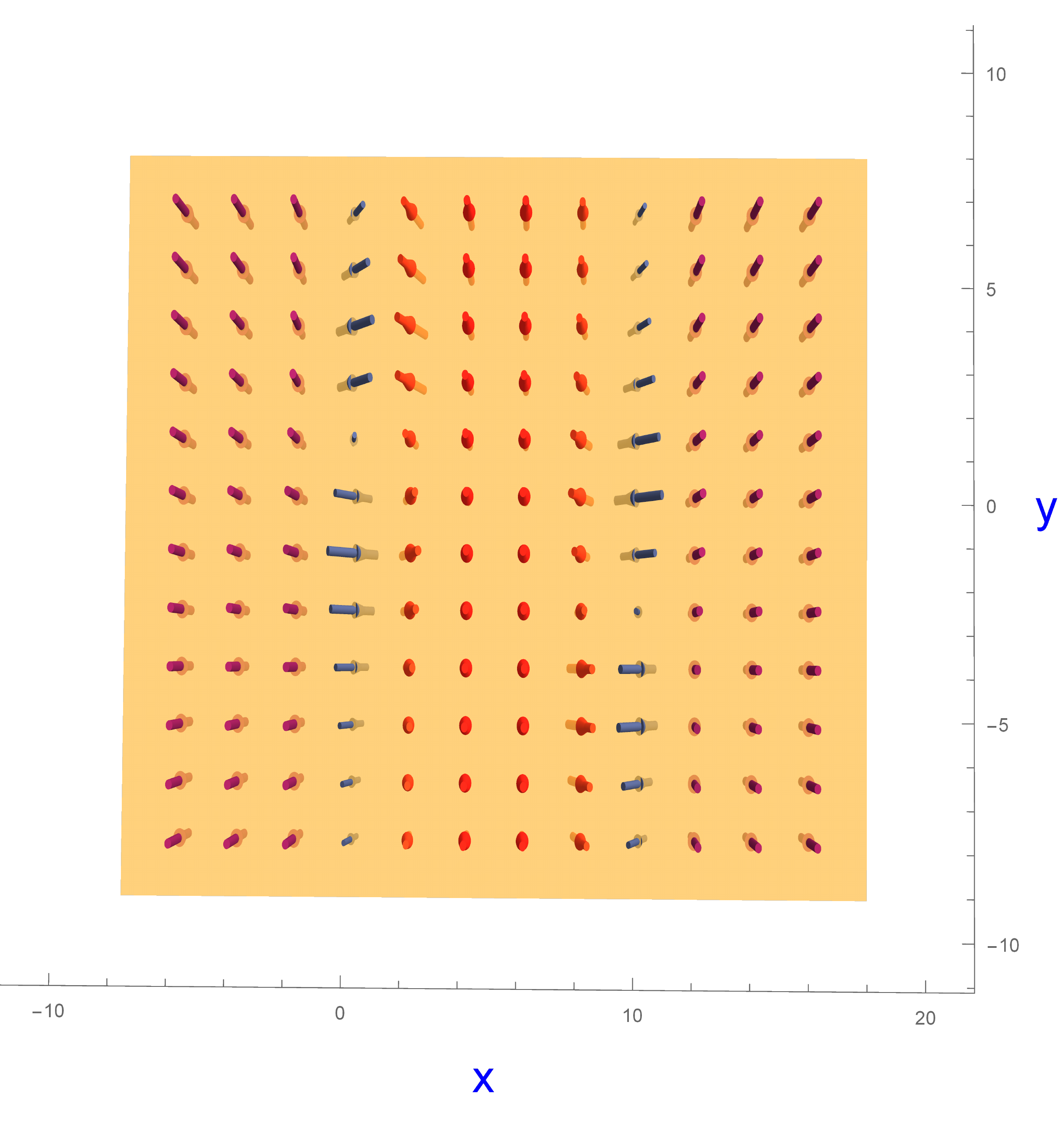}
 \caption{The energy density and magnetisation plots for the domain wall anti-domain wall superposition of \eqref{eq:wall-anti-wall superposition} with $X^{1}=0, X^{2}=10$, $(\varphi_{2})\varphi_{1}$ is a (anti-)kink centred at $x^{2}=\pm2$.}
 \label{fig:superposition plots}
\end{figure}

Turning to the question of how the energy density varies with the $x^{2}$ separation of the kinks we do not get such clear results. The energy density and a magnetisation plot for the case of a kink centred at $x^{2}=-2$ on the domain wall, and an anti-kink centred at $x^{2}=2$ on the anti-domain wall is shown in Fig.~\ref{fig:superposition plots}. As we can vary the relative positions of the centre of the kink along the walls we can compute the energy and find that there is no change. This suggests that the kinks can be thought of as being free to move along the wall. More sophisticated numerical studies are needed to show if they are actually free or if the interaction is just very small.
 
%

\section{\label{sec:conclusions}Summary and Outlook}
In this work, we have studied the effective energy of straight domain walls in chiral magnets and explored its relationship to the sine-Gordon model. This extends the work of Ref.~\cite{Nitta:2012xq,*Kobayashi:2013ju} on domain wall skyrmions by including a DM term. The advantage of this is that the domain wall effective energy becomes the sine-Gordon model without need of the extra anisotropy term used in Refs.~\cite{Nitta:2012xq,*Kobayashi:2013ju}.  
We have found that 
both skyrmions and anti-skyrmions are stable on the domain wall, 
in contrast to the bulk where only either skyrmions or anti-skyrmions are stable 
and the others are unstable.
We also have constructed a domain wall skyrmion lattice.
We then have considered superpositions of domain walls and anti-domain walls with kinks on each of them and found that these kinks are approximately free.

Possible applications to nanodevices will be the most important
future problem. 
For instance, the position of a domain wall can be fixed by 
pinning on impurities. 
Then, the domain wall may play a role of a guide or a rail
for skyrmions. 
The fact that 
both skyrmion and anti-skyrmion are stable on the domain wall 
may be useful. 
As a store of skyrmions, one can reserve either skyrmions or anti-skyrmions 
on the domain wall, which is impossible in the bulk outside the domain wall. 
For instance, if we put both skyrmions and anti-skyrmions, 
they can pair annihilate each other. 
Along the domain wall one can avoid pair annihilation since the sine-Gordon model is known to admit 
a breather solution in which a pair of soliton and anti-soliton oscillate 
without annihilation. 
As for another application, 
a junction of multiple domain walls may be useful.

As already pointed out in Refs.~\cite{Nitta:2012xq,*Kobayashi:2013ju}, 
domain wall skyrmions are equivalent to Josephson vortices 
 in a Josephson junction of two superconductors. 
In this case, the difference between phases of two superconductors 
yields a dynamical degree of freedom which is described by the sine-Gordon model
\cite{USTINOV1998315,*Ustinov2015}. 
These Josephson vortices are nothing but Abrikosov vortices in supercondctors.
Therefore, technologies developed in Josephson vortices 
can be imported to domain wall skyrmions in magnets.

On the more theoretical side, 
the $O(3)$ model is equivalent to the ${\mathbb C}P^1$ model, 
which can be generalized to 
the ${\mathbb C}P^{N-1}$ model, 
admitting ${\mathbb C}P^{N-1}$ skyrmions. 
The ${\mathbb C}P^{N-1}$ model with a potential term 
that is a generalization of the easy-axis potential 
also admits $N$ parallel domain walls 
\cite{Gauntlett:2000ib,*Tong:2002hi}.
With a potential term generalizing the second isotropy term,
${\mathbb C}P^{N-1}$ skyrmions become  
$U(1)^{N-1}$ coupled sine-Gordon solitons inside 
domain walls \cite{Fujimori:2016tmw}. 
On the other hand, 
${\mathbb C}P^{N-1}$ skyrmions 
were also discussed in a
${\mathbb C}P^{N-1}$ model with a generalized DM term 
\cite{Akagi:2021dpk,*Amari:2022boe,*Akagi:2021lva}.
Thus, the 
${\mathbb C}P^{N-1}$ model with the generalized DM term 
and a generalized easy-axis potential should 
admit ${\mathbb C}P^{N-1}$ domain wall skyrmions.
A junction of domain walls can also stably exist
in the ${\mathbb C}P^{N-1}$ model with 
a more general extension of the easy-axis potential 
\cite{Eto:2005cp,*Eto:2005fm,*Eto:2006pg}. 
Domain wall skyrmions on a domain wall junction 
should be useful for nanotechnology.

\begin{acknowledgments}
We would like to thank Norisuke Sakai for useful discussions at an early stage of this work. C.R. thanks Bruno Barton-Singer for general discussions about domain wall skyrmions. The work of M.N.~is supported in part by JSPS Grant-in-Aid for
Scientific Research (KAKENHI Grant No.~JP18H01217 and JP22H01221).
\end{acknowledgments}

\appendix

\section{\label{appendix:kink solutions} Derivation of the kink solutions}
To derive the kink solutions recall that for a one-dimensional problem there is a conserved first integral of motion
\begin{equation}
\left(\frac{d\varphi}{dx^{2}}\right)^{2}-U(\varphi)=C_{0}=\text{constant}.\label{eq:first integral}
\end{equation}
This first integral reduces to the first order equation
\begin{equation}
\frac{d\varphi}{dx^{2}}=\pm\sqrt{U(\varphi)+C_{0}}.
\end{equation}
Here $U(\varphi)=\pm \tilde{\kappa}\left[\sin\left(\varphi+\alpha\right)-\partial_{2}X\cos\left(\varphi+\alpha\right)\right]$ is the potential energy piece of the effective energy. Focusing on the case of a kink on a domain wall  with non-constant translation modulus, $\partial_{2}X=\gamma$: When the translation modulus is linear in $x^{2}$ the potential is
\begin{equation}
\begin{split}
U(\varphi)&=\tilde{\kappa}\left[\sin\left(\varphi+\alpha\right)-\gamma\cos\left(\varphi+\alpha\right)\right]\\
&= \tilde{\kappa}\left(1-\sqrt{1+\gamma^{2}}\sin\left(\varphi+\alpha+\theta-\delta\right)\right)
\end{split},
\end{equation}
with $\tan\delta=\gamma$. If the constant $C_{0}$ is taken to be $C_{0}=\tilde{\kappa}\left(\sqrt{1+\gamma^{2}}-1\right)$ the first order equation becomes
\begin{equation}
\begin{split}
\frac{d\varphi}{dx^{2}}&=\sqrt{\tilde{\kappa}\left(1-\sqrt{1+\gamma^{2}}\sin\left(\varphi+\alpha+\theta-\delta\right)\right)+\tilde{\kappa}\left(\sqrt{1+\gamma^{2}}-1\right)}\\
&=\sqrt{\tilde{\kappa}\sqrt{1+\gamma^{2}}\left(1-\sin\left(\varphi+\alpha+\theta-\delta\right)\right)}
\end{split}.
\end{equation}
This is solved by making the substitution $\Phi=\varphi+\alpha+\theta-\delta-\frac{\pi}{2}$ and then directly integrating
\begin{equation}
\frac{d\Phi}{\sqrt{1-\cos\Phi}} =\pm\sqrt{\tilde{\kappa}\sqrt{1+\gamma^{2}}}dx^{2},
\end{equation}
to give a kink or anti kink depending on the $\pm$ centred at $x^{2}=c$ 
\begin{equation}
\Phi=4\arctan\left(\exp\left(\pm \left(x^{2}-c\right)\sqrt{\frac{\tilde{\kappa}}{2}\sqrt{1+\gamma^{2}}}\right)\right).
\end{equation}
For the anti-domain wall, the only change is that the $\frac{\pi}{2}$ is added rather than subtracted when going from $\varphi$ to $\Phi$. 

\section{\label{appendix:multikink solutions} Derivation of the multikink solutions}
The single kink is not the only solution for the sine-Gordon model, if we return to the first order equation
\begin{equation}
\frac{d\varphi}{dx^{2}}=\pm\sqrt{U(\varphi)+C_{0}},
\end{equation}
then there are more general spiral solutions. To see these take
\begin{equation}
\Phi=\alpha+\theta+\varphi\pm\frac{\pi}{2},
\end{equation}
as above. The first order equation then becomes
\begin{equation}
\frac{d\Phi}{dx^{2}}=\pm\sqrt{C_{0}+\tilde{\kappa}}\sqrt{1-k^{2}\cos^{2}\left(\frac{\Phi}{2}\right)},
\end{equation}
where $k^{2}=\frac{2\tilde{\kappa}}{C_{0}+\tilde{\kappa}}$ is the elliptic modulus. This has a solution in terms of the elliptic Jacobi amplitude as
\begin{equation}
\Phi=\pm\text{am}\left(\frac{\sqrt{C_{0}+\tilde{\kappa}}}{2}x,k\right)-\pi.
\end{equation}

\bibliographystyle{unsrt}
\bibliography{dw_skyrmion}

\begin{thebibliography}{10}

\bibitem{Skyrme:1962vh}
T.~H.~R. Skyrme.
\newblock {A Unified Field Theory of Mesons and Baryons}.
\newblock {\em Nucl. Phys.}, 31:556--569, 1962.

\bibitem{Witten:1983tx}
E.~Witten.
\newblock {Current Algebra, Baryons, and Quark Confinement}.
\newblock {\em Nucl. Phys. B}, 223:433--444, 1983.

\bibitem{MantonsBook}
N.S. Manton.
\newblock {\em {Skyrmions - A Theory of Nuclei}}.
\newblock World Scientific, Singapore, 2022.

\bibitem{multifaceted}
M.~Rho and I.~Zahed, editors.
\newblock {\em {The Multifaceted Skyrmions}}.
\newblock World Scientific, Singapore, second edition, 2016.

\bibitem{Manton:2004tk}
N.~S. Manton and P.~M. Sutcliffe.
\newblock {\em Topological solitons.}
\newblock Cambridge Monographs on Mathematical Physics. Cambridge University
  Press, Cambridge, July 2004.

\bibitem{Bogdanov:1989}
A.N. Bogdanov and D.A. Yablonskii.
\newblock {Thermodynamically stable vortices in magnetically ordered crystals.
  The mixed state of magnets}.
\newblock {\em Sov. Phys. JETP}, 68:101--103, 1989.

\bibitem{Bogdanov:1995}
A.~Bogdanov.
\newblock {New localized solutions of the nonlinear field equations}.
\newblock {\em JETP Lett.}, 62:247--251, 1995.

\bibitem{Dzyaloshinskii}
I.~Dzyaloshinskii.
\newblock {A Thermodynamic Theory of `Weak' Ferromagnetism of
  Antiferromagnetics}.
\newblock {\em J.~Phys.~Chem.~Solids}, 4:241--255, 1958.

\bibitem{Moriya:1960zz}
T.~Moriya.
\newblock {Anisotropic Superexchange Interaction and Weak Ferromagnetism}.
\newblock {\em Phys. Rev.}, 120:91--98, 1960.

\bibitem{doi:10.1126/science.1166767}
S.~M\"{u}hlbauer, B.~Binz, F.~Jonietz, C.~Pfleiderer, A.~Rosch, A.~Neubauer,
  R.~Georgii, and P.~Boni.
\newblock Skyrmion lattice in a chiral magnet.
\newblock {\em Science}, 323(5916):915--919, 2009.

\bibitem{doi:10.1038/nature09124}
X.~Z. Yu, Y.~Onose, N.~Kanazawa, J.~H. Park, J.~H. Han, Y.~Matsui, N.~Nagaosa,
  and Y.~Tokura.
\newblock {Real-space observation of a two-dimensional skyrmion crystal}.
\newblock {\em Nature}, 465:901--904, 06 2010.

\bibitem{doi:10.1038/nphys2045}
S.~Heinze, K.~von Bergmann, M.~Menzel, J.~Brede, A.~Kubetzka, R.~Wiesendanger,
  G.~Bihlmayer, and S.~Blugel.
\newblock {Spontaneous atomic-scale magnetic skyrmion lattice in two
  dimensions}.
\newblock {\em Nature Physics}, 7:713--718, 09 2011.

\bibitem{doi:10.1038/nnano.2013.29}
A.~Fert, V.~Cros, and J.~Sampaio.
\newblock {Skyrmions on the track}.
\newblock {\em Nature Nanotechnology}, 8:152--156, 03 2013.

\bibitem{togawa2012chiral}
Y.~Togawa, T.~Koyama, K.~Takayanagi, S.~Mori, Y.~Kousaka, J.~Akimitsu,
  S.~Nishihara, K.~Inoue, A.~S. Ovchinnikov, and J-i. Kishine.
\newblock Chiral magnetic soliton lattice on a chiral helimagnet.
\newblock {\em Phys.Rev.Lett.}, 108(10):107202, 2012.

\bibitem{togawa2016symmetry}
Y.~Togawa, Y.~Kousaka, K.~Inoue, and J-i. Kishine.
\newblock Symmetry, structure, and dynamics of monoaxial chiral magnets.
\newblock {\em Journal of the Physical Society of Japan}, 85(11):112001, 2016.

\bibitem{KISHINE20151}
J-i. Kishine and A.~S. Ovchinnikov.
\newblock Chapter one - theory of monoaxial chiral helimagnet.
\newblock volume~66 of {\em Solid State Physics}, pages 1--130. Academic Press,
  2015.

\bibitem{PhysRevB.97.184303}
A.~A. Tereshchenko, A.~S. Ovchinnikov, I.~Proskurin, E.~V. Sinitsyn, and J-i.
  Kishine.
\newblock Theory of magnetoelastic resonance in a monoaxial chiral helimagnet.
\newblock {\em Phys. Rev. B}, 97:184303, May 2018.

\bibitem{PhysRevB.65.064433}
J.~Chovan, N.~Papanicolaou, and S.~Komineas.
\newblock Intermediate phase in the spiral antiferromagnet
  ${\mathrm{ba}}_{2}{\mathrm{cuge}}_{2}{\mathrm{o}}_{7}$.
\newblock {\em Phys. Rev. B}, 65:064433, Jan 2002.

\bibitem{Ross:2020orc}
C.~Ross and M.~Sakai, N.and~Nitta.
\newblock {Exact ground states and domain walls in one dimensional chiral
  magnets}.
\newblock {\em JHEP}, 12:163, 2021.

\bibitem{Rossler:2006}
U.~K. Rossler, A.~N. Bogdanov, and C.~Pfleiderer.
\newblock Spontaneous skyrmion ground states in magnetic metals.
\newblock {\em Nature}, 442:797, 2006.

\bibitem{Han:2010by}
J.~H. Han, J.~Zang, Z.~Yang, J-H. Park, and N.~Nagaosa.
\newblock {Skyrmion Lattice in Two-Dimensional Chiral Magnet}.
\newblock {\em Phys. Rev. B}, 82:094429, 2010.

\bibitem{Lin:2014ada}
S-Z. Lin, A.~Saxena, and C.~D. Batista.
\newblock {Skyrmion fractionalization and merons in chiral magnets with
  easy-plane anisotropy}.
\newblock {\em Phys. Rev. B}, 91(22):224407, 2015.

\bibitem{Ross:2020hsw}
C.~Ross, N.~Sakai, and M.~Nitta.
\newblock {Skyrmion interactions and lattices in chiral magnets: analytical
  results}.
\newblock {\em JHEP}, 02:095, 2021.

\bibitem{Romming:2013}
N.~Romming, C.~Hanneken, M.~Menzel, J.~E. Bickel, B.~Wolter, K.~von Bergmann,
  A.~Kubetzka, and R.~Wiesendanger.
\newblock {Writing and Deleting Single Magnetic Skyrmions}.
\newblock {\em Science}, 341:636--639, 2013.

\bibitem{Wolf:2022}
{D.~Wolf, S. Schneider, U.K. R\"o\ss ler, A. Kov{\'a}cs, M. Schmidt, R.E.
  Dunin-Borkowski, B. B{\"u}chner, B.Rellinghaus, and A. Lubk}.
\newblock Unveiling the three-dimensional magnetic texture of skyrmion tubes.
\newblock {\em Nature Nanotechnology}, 17:250--255, 2022.

\bibitem{doi:10.1126/science.1145799}
S.~S.~P. Parkin, M.~Hayashi, and L.Thomas.
\newblock Magnetic domain-wall racetrack memory.
\newblock {\em Science}, 320(5873):190--194, 2008.

\bibitem{KUMAR20221}
D.~Kumar, T.~Jin, R.Sbiaa, M.~Kl{\:a}ui, S.~Bedanta, S.~Fukami, D.~Ravelosona,
  S-H. Yang, X.~Liu, and S.N. Piramanayagam.
\newblock Domain wall memory: Physics, materials, and devices.
\newblock {\em Physics Reports}, 958:1--35, 2022.
\newblock Domain Wall Memory: Physics, Materials, and Devices.

\bibitem{tomasello2014strategy}
R.~Tomasello, E.~Martinez, R.~Zivieri, L.~Torres, M.~Carpentieri, and
  G.~Finocchio.
\newblock A strategy for the design of skyrmion racetrack memories.
\newblock {\em Scientific reports}, 4(1):1--7, 2014.

\bibitem{Nitta:2012xq}
M.~Nitta.
\newblock {Josephson vortices and the Atiyah-Manton construction}.
\newblock {\em Phys. Rev. D}, 86:125004, 2012.

\bibitem{Kobayashi:2013ju}
M.~Kobayashi and M.~Nitta.
\newblock {Sine-Gordon kinks on a domain wall ring}.
\newblock {\em Phys. Rev. D}, 87(8):085003, 2013.

\bibitem{Eto:2005cc}
M.~Eto, M.~Nitta, K.~Ohashi, and D.~Tong.
\newblock {Skyrmions from instantons inside domain walls}.
\newblock {\em Phys. Rev. Lett.}, 95:252003, 2005.

\bibitem{PhysRevB.102.094402}
S.~Lepadatu.
\newblock Emergence of transient domain wall skyrmions after ultrafast
  demagnetization.
\newblock {\em Phys. Rev. B}, 102:094402, Sep 2020.

\bibitem{Nagase:2020imn}
T.Nagase, Y-G. So, H.~Yasui, T.~Ishida, H.~K. Yoshida, Y.~Tanaka, K.~Saitoh,
  N.~Ikarashi, Y.~Kawaguchi, M.~Kuwahara, and M.~Nagao.
\newblock {Observation of domain wall bimerons in chiral magnets}.
\newblock {\em Nature Commun.}, 12(1):3490, 2021.

\bibitem{Yang:2021}
K.~Yang, K.~Nagase, and Y.~Hirayama~et.al.
\newblock {Wigner solids of domain wall skyrmions}.
\newblock {\em Nat Commun}, 12:6006, 2021.

\bibitem{Kim:2017lsi}
S.~K. Kim and Y.~Tserkovnyak.
\newblock {Magnetic Domain Walls as Hosts of Spin Superfluids and Generators of
  Skyrmions}.
\newblock {\em Phys. Rev. Lett.}, 119(4):047202, 2017.

\bibitem{PhysRevB.99.184412}
R.~Cheng, M.~Li, A.~Sapkota, A.~Rai, A.~Pokhrel, T.~Mewes, C.~Mewes, D.~Xiao,
  M.~De~Graef, and V.~Sokalski.
\newblock Magnetic domain wall skyrmions.
\newblock {\em Phys. Rev. B}, 99:184412, May 2019.

\bibitem{KBRBSK}
V.~M. Kuchkin, B.~Barton-Singer, F.~N. Rybakov, S.~Bl\"ugel, B.~J. Schroers,
  and N.~S. Kiselev.
\newblock {Magnetic skyrmions, chiral kinks and holomorphic functions}.
\newblock {\em Phys. Rev. B}, 102(14):144422, 2020.

\bibitem{Jennings:2013aea}
P.~Jennings and P.~Sutcliffe.
\newblock {The dynamics of domain wall Skyrmions}.
\newblock {\em J. Phys. A}, 46:465401, 2013.

\bibitem{Bychkov:2016cwc}
V.~Bychkov, M.~Kreshchuk, and E.~Kurianovych.
\newblock {Strings and skyrmions on domain walls}.
\newblock {\em Int. J. Mod. Phys. A}, 33(18n19):1850111, 2018.

\bibitem{Sutcliffe:1992he}
P.~M. Sutcliffe.
\newblock {Sine-Gordon solitions from CP(1) instantons}.
\newblock {\em Phys. Lett. B}, 283:85--89, 1992.

\bibitem{Stratopoulos:1992hq}
G.~N. Stratopoulos and W.~J. Zakrzewski.
\newblock {Approximate Sine-Gordon solitons}.
\newblock {\em Z. Phys. C}, 59:307--312, 1993.

\bibitem{Kudryavtsev:1997nw}
A.~E. Kudryavtsev, B.~M. A.~G. Piette, and W.~J. Zakrzewski.
\newblock {Skyrmions and domain walls in (2+1)-dimensions}.
\newblock {\em Nonlinearity}, 11:783--795, 1998.

\bibitem{Auzzi:2006ju}
R.~Auzzi, M.~Shifman, and A.~Yung.
\newblock {Domain Lines as Fractional Strings}.
\newblock {\em Phys. Rev. D}, 74:045007, 2006.

\bibitem{Abraham:1992vb}
E.~R.~C. Abraham and P.~K. Townsend.
\newblock {Q kinks}.
\newblock {\em Phys. Lett. B}, 291:85--88, 1992.

\bibitem{Abraham:1992qv}
E.~R.~C. Abraham and P.~K. Townsend.
\newblock {More on Q kinks: A (1+1)-dimensional analog of dyons}.
\newblock {\em Phys. Lett. B}, 295:225--232, 1992.

\bibitem{Arai:2002xa}
M.~Arai, M.~Naganuma, M.~Nitta, and N.~Sakai.
\newblock {Manifest supersymmetry for BPS walls in N=2 nonlinear sigma models}.
\newblock {\em Nucl. Phys. B}, 652:35--71, 2003.

\bibitem{Arai:2003es}
M.~Arai, M.~Naganuma, M.~Nitta, and N.~Sakai.
\newblock {BPS wall in N=2 SUSY nonlinear sigma model with Eguchi-Hanson
  manifold}.
\newblock pages 299--325, 2 2003.

\bibitem{Manton:1981mp}
N.~S. Manton.
\newblock {A Remark on the Scattering of BPS Monopoles}.
\newblock {\em Phys. Lett. B}, 110:54--56, 1982.

\bibitem{Eto:2006uw}
M.~Eto, Y.~Isozumi, M.~Nitta, K.~Ohashi, and N.~Sakai.
\newblock {Manifestly supersymmetric effective Lagrangians on BPS solitons}.
\newblock {\em Phys. Rev. D}, 73:125008, 2006.

\bibitem{Fujimori:2016tmw}
T.~Fujimori, H.~Iida, and M.~Nitta.
\newblock {Field theoretical model of multilayered Josephson junction and
  dynamics of Josephson vortices}.
\newblock {\em Phys. Rev. B}, 94(10):104504, 2016.

\bibitem{Gauntlett:2000ib}
J.~P. Gauntlett, D.~Tong, and P.~K. Townsend.
\newblock {Multidomain walls in massive supersymmetric sigma models}.
\newblock {\em Phys. Rev. D}, 64:025010, 2001.

\bibitem{Tong:2002hi}
D.~Tong.
\newblock {The Moduli space of BPS domain walls}.
\newblock {\em Phys. Rev. D}, 66:025013, 2002.

\bibitem{Nitta:2015mma}
M.~Nitta.
\newblock {Josephson junction of non-Abelian superconductors and non-Abelian
  Josephson vortices}.
\newblock {\em Nucl. Phys. B}, 899:78--90, 2015.

\bibitem{Nitta:2015mxa}
M.~Nitta.
\newblock {Josephson instantons and Josephson monopoles in a non-Abelian
  Josephson junction}.
\newblock {\em Phys. Rev. D}, 92(4):045010, 2015.

\bibitem{Nitta:2014rxa}
M.~Nitta.
\newblock {Non-Abelian Sine-Gordon Solitons}.
\newblock {\em Nucl. Phys. B}, 895:288--302, 2015.

\bibitem{Eto:2015uqa}
M.~Eto and M.~Nitta.
\newblock {Non-Abelian Sine-Gordon Solitons: Correspondence between $SU(N)$
  Skyrmions and ${\mathbb C}P^{N-1}$ Lumps}.
\newblock {\em Phys. Rev. D}, 91(8):085044, 2015.

\bibitem{Shifman:2003uh}
M.~Shifman and A.~Yung.
\newblock {Localization of nonAbelian gauge fields on domain walls at weak
  coupling (D-brane prototypes II)}.
\newblock {\em Phys. Rev. D}, 70:025013, 2004.

\bibitem{Eto:2008dm}
M.~Eto, T.~Fujimori, M.~Nitta, K.~Ohashi, and N.~Sakai.
\newblock {Domain walls with non-Abelian clouds}.
\newblock {\em Phys. Rev. D}, 77:125008, 2008.

\bibitem{Nitta:2012wi}
M.~Nitta.
\newblock {Correspondence between Skyrmions in 2+1 and 3+1 Dimensions}.
\newblock {\em Phys. Rev. D}, 87(2):025013, 2013.

\bibitem{Gudnason:2014nba}
S.~B. Gudnason and M.~Nitta.
\newblock {Domain wall Skyrmions}.
\newblock {\em Phys. Rev. D}, 89(8):085022, 2014.

\bibitem{Gudnason:2014hsa}
S.~B. Gudnason and M.~Nitta.
\newblock {Incarnations of Skyrmions}.
\newblock {\em Phys. Rev. D}, 90(8):085007, 2014.

\bibitem{Kudryavtsev:1999zm}
A.~E. Kudryavtsev, B.~M. A.~G. Piette, and W.~J. Zakrzewski.
\newblock {On the interactions of skyrmions with domain walls}.
\newblock {\em Phys. Rev. D}, 61:025016, 2000.

\bibitem{Gudnason:2013qba}
S.~B. Gudnason and M.~Nitta.
\newblock {Baryonic sphere: a spherical domain wall carrying baryon number}.
\newblock {\em Phys. Rev. D}, 89(2):025012, 2014.

\bibitem{Gudnason:2018oyx}
S.~B. Gudnason and M.~Nitta.
\newblock {Baryonic handles: Skyrmions as open vortex strings on a domain
  wall}.
\newblock {\em Phys. Rev. D}, 98(12):125002, 2018.

\bibitem{Nitta:2012rq}
M.~Nitta.
\newblock {Matryoshka Skyrmions}.
\newblock {\em Nucl. Phys. B}, 872:62--71, 2013.

\bibitem{Nitta:2013cn}
M.~Nitta.
\newblock {Instantons confined by monopole strings}.
\newblock {\em Phys. Rev. D}, 87(6):066008, 2013.

\bibitem{Nitta:2013vaa}
M.~Nitta.
\newblock {Incarnations of Instantons}.
\newblock {\em Nucl. Phys. B}, 885:493--504, 2014.

\bibitem{Nitta:2022ahj}
M.~Nitta.
\newblock {Relations among topological solitons}.
\newblock 2 2022.

\bibitem{Derrick:1964}
G.~H. Derrick.
\newblock Comments on nonlinear wave equations as models for elementary
  particles.
\newblock {\em J. Math. Phys.}, 5:1252–1254, 1964.

\bibitem{BRS}
B.~Barton-Singer, C.~Ross, and B.~J. Schroers.
\newblock {Magnetic Skyrmions at Critical Coupling}.
\newblock {\em Commun. Math. Phys.}, 375(3):2259--2280, 2020.

\bibitem{Nitta:2012kj}
M.~Nitta.
\newblock {Defect formation from defect--anti-defect annihilations}.
\newblock {\em Phys. Rev. D}, 85:101702, 2012.

\bibitem{Nitta:2012kk}
M.~Nitta.
\newblock {Knots from wall--anti-wall annihilations with stretched strings}.
\newblock {\em Phys. Rev. D}, 85:121701, 2012.

\bibitem{Nitta:2012mg}
M.~Nitta.
\newblock {Knotted instantons from annihilations of monopole-instanton
  complex}.
\newblock {\em Int. J. Mod. Phys. A}, 28:1350172, 2013.

\bibitem{USTINOV1998315}
A.V. Ustinov.
\newblock Solitons in josephson junctions.
\newblock {\em Physica D: Nonlinear Phenomena}, 123(1):315--329, 1998.
\newblock Annual International Conference of the Center for Nonlinear Studies.

\bibitem{Ustinov2015}
A.~V. Ustinov.
\newblock {\em Solitons in Josephson Junctions: Physics of Magnetic Fluxons in
  Superconducting Junctions and Arrays}.
\newblock Wiley-VCH, 2015.

\bibitem{Akagi:2021dpk}
Y.~Akagi, Y.~Amari, N.~Sawado, and Y.~Shnir.
\newblock {Isolated skyrmions in the $CP^2$ nonlinear sigma model with a
  Dzyaloshinskii-Moriya type interaction}.
\newblock {\em Phys. Rev. D}, 103(6):065008, 2021.

\bibitem{Amari:2022boe}
Y.~Amari, Y.~Akagi, S.~B. Gudnason, M.~Nitta, and Y.~Shnir.
\newblock {$\mathbb{C}P^2$ Skyrmion Crystals in an SU(3) Magnet with a
  Generalized Dzyaloshinskii-Moriya Interaction}.
\newblock 4 2022.

\bibitem{Akagi:2021lva}
Y.~Akagi, Y.~Amari, S.~B. Gudnason, M.~Nitta, and Y.~Shnir.
\newblock {Fractional Skyrmion molecules in a \ensuremath{\mathbb{C}}P$^{N-1}$
  model}.
\newblock {\em JHEP}, 11:194, 2021.

\bibitem{Eto:2005cp}
M.~Eto, Y.~Isozumi, M.~Nitta, K.~Ohashi, and N.~Sakai.
\newblock {Webs of walls}.
\newblock {\em Phys. Rev. D}, 72:085004, 2005.

\bibitem{Eto:2005fm}
M.~Eto, Y.~Isozumi, M.~Nitta, K.~Ohashi, and N.~Sakai.
\newblock {Non-Abelian webs of walls}.
\newblock {\em Phys. Lett. B}, 632:384--392, 2006.

\bibitem{Eto:2006pg}
M.~Eto, Y.~Isozumi, M.~Nitta, K.~Ohashi, and N.~Sakai.
\newblock {Solitons in the Higgs phase: The Moduli matrix approach}.
\newblock {\em J. Phys. A}, 39:R315--R392, 2006.

\end{thebibliography}

\end{document}